# Comparison of the use of NiFe and CoFe as electrodes for metallic lateral spin-valves

G. Zahnd[1,2], L. Vila[1,2,*], T. V. Pham[1,2], A. Marty[1,2], P. Laczkowski[1,2], W. Savero Torres[1,2], C. Beigné[1,2], C. Vergnaud[1,2], M. Jamet[1,2] and J.-P. Attané[1,2,†]

[1]Université Grenoble Alpes, INAC-SP2M, F-38000 Grenoble, France
[2]CEA, Institut Nanosciences et Cryogénie, SP2M, F-38000 Grenoble, France

**Abstract:**
**Spin injection and detection in $Co_{60}Fe_{40}$-based all-metallic lateral spin-valves have been studied at both room and low temperatures. The obtained spin signals amplitudes have been compared to that of identical $Ni_{80}Fe_{20}$-based devices. The replacement of $Ni_{80}Fe_{20}$ by CoFe allows increasing the spin signal amplitude by up to one order of magnitude, thus reaching 50 mΩ at room temperature. The spin signal dependence with the distance between the ferromagnetic electrodes has been analyzed using both a 1D spin transport model and finite elements method simulations. The enhancement of the spin signal amplitude when using CoFe electrodes can be explained by a higher effective polarization.**

I. INTRODUCTION
Lateral spin valves (LSVs) are nowadays widely used in spintronics. Non-local measurements in LSVs are one of the main tool to study of spin injection, spin transport [1] and precession [2] in non-magnetic material [3,4] or through interfaces [5,6]. LSVs also enable studies of spin-orbitronics effects, as non-local measurements allow separating the spin-orbit material from the ferromagnetic one [7,8,9,10]. The ability to generate and detect pure spin currents in these devices provide ways to study spin-calorimetry [11,12], and spin transfer [13,14] by pure spin currents. With the exception of a recent work on CoPd and CoNi-based LSVs [15], all lateral have been made using soft magnetic materials with in-plane magnetization. The Permalloy ($Ni_{80}Fe_{20}$ alloy) is currently the most employed ferromagnet in LSVs [3,4,5,6,7,8,9,10,11,12,13,14,16,17,18,19,20,21]. This is due to its relatively high polarization, its simple deposition method on most materials, and to the fact that its magnetization is easily controlled by external fields. Also, as Permalloy saturation magnetization is quite low, it is a good candidate to realize spin transfer torque experiments. Nevertheless, NiFe-based LSVs emit small signals, typically of a few milliohms. This is an important limitation, obviously for applications such as hard disk-drive read-heads [22], but also for fundamental studies: for instance, studies concerning spin precession within the channel [16,19], spin transfer torque [13,14] or spin calorimetrics [11,12] require highly efficient LSVs.

---

[*] Laurent.vila@cea.fr
[†] Jean-philippe.attane@cea.fr

Recently, Heuslers alloys with high polarization have shown to lead to consequent output voltages [22,23,24]. However, as they require epitaxial growth, these materials are relatively difficult to handle, which might restrain their use to fundamental research. Other in-plane magnetic materials have also been used as LSVs electrodes, such as Co [2], but their spin signal amplitudes did not match the results obtained with NiFe. Ni-based LSVs have not shown sizeable signals [25]. The use of barriers allows increasing significantly the spin signal amplitude by solving the spin resistance mismatch between the non-ferromagnetic and the ferromagnetic metals [19], but necessarily increases the resistance of the device. Hence, replacing the usual NiFe by a versatile material that could lead to high signal amplitudes remains a challenging problem for both fundamental studies and applications.

CoFe alloys have recently been studied in LSVs, in order to highlight the interest of CoFeAl Heuslers alloys [23,24]. Although the spin signals obtained with CoFe remained small (the widths of the devices being larger than 100 nm), the measured polarization of the CoFe electrode was found to be relatively large (0.45 at 77K) [23], and CoFe could therefore be a good candidate for the replacement of NiFe.

In this paper, we develop a systematic comparison of CoFe-based and NiFe-based LSVs. Using CoFe, non-local spin signals as large as 50 mΩ and 80 mΩ were obtained at 300 K and 10 K respectively. We show that the replacement of $Co_{60}Fe_{40}$ by $Ni_{80}Fe_{20}$ leads to an increase of the signal of one order of magnitude. We detail the fabrication process, and compare NiFe and CoFe-based LSVs with identical geometries, for channels of both Cu and Al. CoFe spin-transport parameters are then extracted by using a 1D spin transport model, whose validity is checked by finite element method simulation.

II. EXPERIMENTAL DETAILS
Sets of LSVs have been patterned by e-beam lithography on a $SiO_2$ substrate, with gaps (*i.e.,* electrode-electrode distance from center to center) varying from 150 to 1500 nm (cf. fig. 1(a) and 1(b)). The ferromagnetic nanowires have been fabricated by evaporation of $Co_{60}Fe_{40}$ pellets through a patterned PMMA resist mask and subsequent lift-off. The non-magnetic wires have been made in a second lithography step. An Argon Ion Beam milling has been used in order to obtain clean transparent interfaces between the non-magnetic channel and the ferromagnetic electrodes. Both Cu and Al have been used as non-magnetic materials, in two different sets of LSVs. Each wire is nominally 50 nm wide, CoFe nanowires are 15 nm thick, and nonmagnetic wires 80 nm thick. Figure 1(a) shows the probe configuration for non-local measurements. These geometries have been chosen to correspond to previously measured NiFe-based LSVs [20]. As spin signal amplitudes strongly depend on the geometries, the materials and the interfaces qualities, it is important to mention that all our LSVs have been patterned using the same process, and that they are geometrically identical.

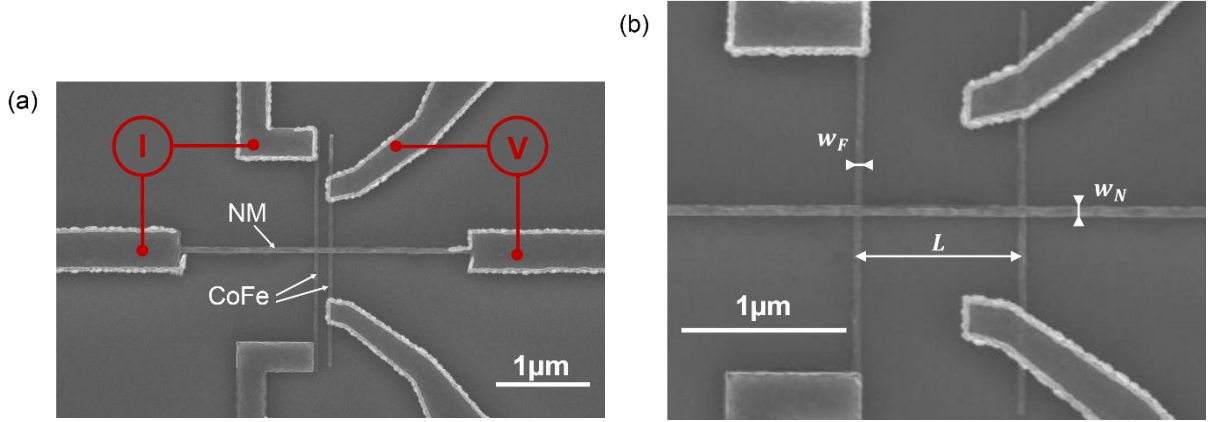

**Figure 1**
a) SEM image of a 150 nm gap lateral spin valve (LSV), and electrical scheme for non-local measurement. b) SEM image of an identical kind of structure, with a gap of 1µm. The CoFe electrodes have sections of 50*15 nm², and the magnetic material (here Al) canal is nominally 50 nm wide and 80 nm thick.

III. RESULTS

Figure 2(a) presents non-local measurements at 300 K of LSVs with 150 nm-long gaps, for both CoFe/Al and NiFe/Al devices. For each curve, the two voltage levels correspond to the parallel and antiparallel states of magnetization of the ferromagnetic electrodes [26,27]. The drop of spin signal corresponds to the switching of a first magnetic wire, leading to an antiparallel state, and the return to the upper value corresponds to the switching of the second wire, bringing the system back to the parallel state. The slight asymmetry in field that can be observed for the NiFe sample is due to the stochasticity of the magnetization switching process [28]. To obtain different switching fields, we added to one electrode a nucleation pad, whereas the other one is a straight wire. The spin signal amplitude is the difference of voltage detected between the two states, and is of 54.5 mΩ for the CoFe-based LSV, and of 5.4 mΩ for the NiFe-based LSV.

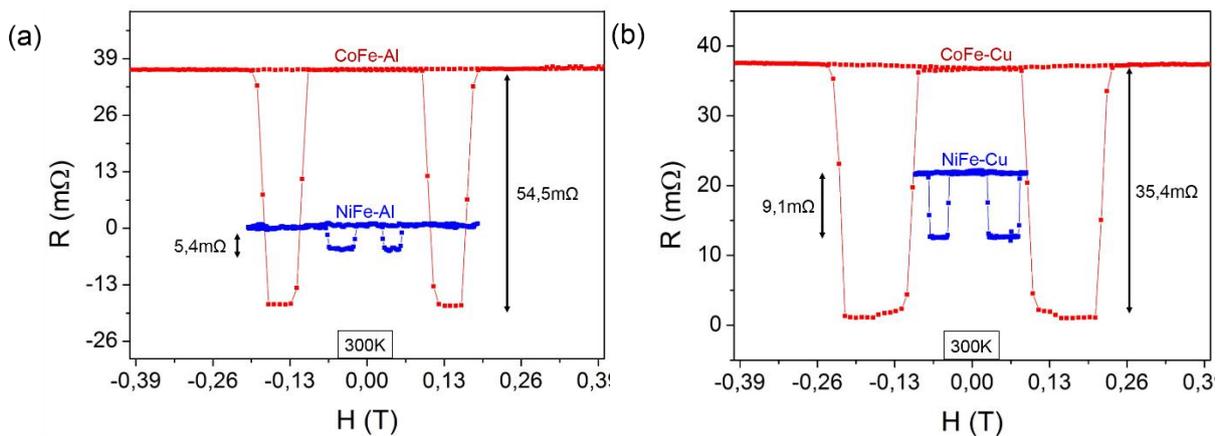

**Figure 2**
a) Non-local measurements at room temperature, for similar CoFe-Al and NiFe-Al LSVs. The gap is 150 nm wide. The spin signal amplitude is 54 mΩ for CoFe, *i.e.* an exact ten fold increase of signal. b) Identical measurement performed on Cu-based LSVs. The spin signal amplitude is 35 mΩ for CoFe, *i.e.* an approximate four fold increase.

On figure 2(b), identical measurements are displayed for Cu-based LSVs, exhibiting respective signals of 35.4mΩ and 9.1mΩ. The replacement of NiFe by CoFe allows obtaining spin signal amplitudes four times higher for the Cu channel, and up to of one order of amplitude higher for the Al channel. At 10K, the spin signal amplitudes for those Cu and Al-based LSVs reach 62.8mΩ and 83.5mΩ respectively.

The spin resistance mismatch is expected to be bigger for Aluminum than Cooper, which possesses smaller spin diffusion length and resistivity, as presented below. Nevertheless, when comparing CoFe-Al and CoFe-Cu LSVs measurements, one can notice that devices with aluminum give higher spin signals, and that it is not the case when using NiFe electrodes. This could be explained by interface effects: the quality of the interface, which can be depend on materials inter-diffusion, is known to have a considerable influence on spin memory loss [29].

IV. ANALYSIS AND DISCUSSION

In order to determine the effective polarization of the CoFe electrodes, devices having different gaps were measured for both Cu and Al channels. The variations of the spin signal amplitudes are shown in figures 3(a) and 3(b). For a given set of LSVs, the hereinafter one-dimensional model [1,30] has been used to analyze the experimental data, with the assumption that interfaces are transparent:

$$\Delta R = \frac{4R_N \left(\frac{P_F}{1-P_F^2} \frac{R_F}{R_N}\right)^2 \times e^{-L/\lambda_N}}{\left(1 + \frac{2}{1-P_F^2} \frac{R_F}{R_N}\right)^2 - e^{-2L/\lambda_N}} \quad (1)$$

Here $\lambda_F$ and $\lambda_N$ are the spin diffusion length of the CoFe and the non-magnetic material, $R_F$ and $R_N$ are their respective spin resistances defined by $R_i = \rho_i \lambda_i / A_i$, $i \in \{N, F\}$, where $A_i$ is the effective cross area of the $i$ material. $P_F$ is the effective polarization, and $L$ is the gap between the injector and detector. From four-probe measurements we extracted material resistivities; $\rho_F^{CoFe}$=28μΩ.cm, $\rho_F^{NiFe}$=23μΩ.cm, $\rho_N^{Al}$=6.3μΩ.cm and $\rho_N^{Cu}$=4.6μΩ.cm at RT. The ferromagnetic materials spin diffusion lengths are difficult to estimate, and our experiment cannot determine $\lambda_F$ and $P_F$ precisely and independently. Further experiments, as spin absorption measurements, should be done to determine them separately. Assuming $\lambda_F \times \rho_F$ independent of the temperature and using the values measured by Ahn *et al.* [31], the spin diffusion length of CoFe in our system can be assumed to lie in the range 2.4-6.5 nm. For the sake of simplicity and to allow comparison with the NiFe systems, the spin diffusion length of CoFe is taken to be similar to that of NiFe (3.5nm). The larger resistivity of CoFe over NiFe induces a better spin resistance matching between the normal material and the ferromagnetic one. However the consequent increase of the spin signal cannot be explained entirely by this spin resistance difference alone.

The fit of our data by the 1D model allows us to extract the values of $\lambda_N$ and $P_F$. Those values are presented in table 1, together with the values of NiFe-based LSVs previously measured.

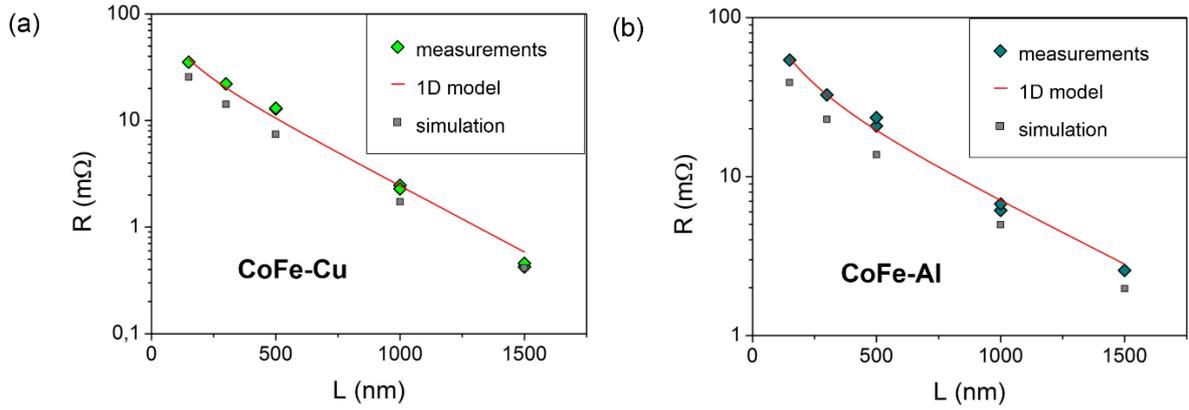

**Figure 3**

a) Room temperature gap dependence realized on CoFe-Cu LSVs and represented by green diamonds on a semi-log scale. The one dimensional model fit is also displayed by the red curve. FEM simulations outputs using the one dimensional model's parameters are eventually plotted (grey squares). b) Identical plot for CoFe-Al LSVs. Measurement values are represented by blues diamonds, the one-dimensional model fit by the red curve, and the FEM simulations outputs in by the grey squares

| Ferromagnetic | Non-magnetic | $\lambda_N$ (nm) | $P_F$ (ø) |
|---|---|---|---|
| CoFe | Cu | 350±80 | 0.50±0.20 |
|  | Al | 550±130 | 0.58±0.20 |
| NiFe | Cu | 290±70 | 0.17±0.05 |
|  | Al | 450±100 | 0.26±0.05 |

*Table 1: Values of the spin diffusion lengths of the normal materials and of the effective polarizations, extracted using a fit of the experimental data at 300K by the 1D model. The relatively large error bars on the effective polarization of CoFe systems take into account the large uncertainty on the value of the spin diffusion length of CoFe discussed previously.*

The significant increase of signal is thus explained by an effective polarization much larger for CoFe devices. Similar spin diffusion lengths of copper and aluminum are obtained when changing from NiFe to CoFe, and are comparable to what can be found in the literature [4,17]. The CoFe polarization is consistent for the two sets of LSVs.

Our analysis assumes that the interfaces are transparent. That supposition has been put to the test by mean of a four point measurement of the interface. Due to the 3D character of the current lines in the cross volume, a finite element method simulation was undertaken. In addition to the bulk resistivity of the wires, an interfacial resistance was added, in order to determine the relationship between the measured voltage and this interfacial resistance [32]. In the given four probes geometry, we measured resistances value smaller than 200 mΩ, which lead to an interface resistance smaller than 1fΩ.m$^{-2}$, thus justifying the transparent interface hypothesis.

A second questioning fact is the one-dimension model validity. Indeed, scales are here considerably reduced, and the geometry for short gaps might not be assimilated as one-dimensional anymore. This possible deviation for the shortest gaps could lead to erroneous values of $\lambda_N$ and $P_F$. Thus, in order to comfort our analysis, we performed finite element method simulations of the non-local injection, based on a two spin-current drift-diffusion model [33]. The simulation has been performed using GMSH [34] for the geometrical construction, the meshing, and the post-processing part, and GETDP [35] as its associated solver (See ref. [21] for details on the simulation method). The material parameters used in these simulations are those given previously. Magnetization has been set along the easy axis of the ferromagnetic wires. Parallel and antiparallel states have been simulated to obtain the spin signal amplitude. A 150 nm gap LSV geometry simulation is presented in fig. 4 in the parallel state case. The spin signal amplitude has been computed for each gap, and both for CoFe-Cu and CoFe-Al LSVs. These values are displayed on figures 3(a) and 3(b). As expected, the simulated spin signal amplitudes are in good agreement with the 1D model, which suggest that the 1D model is valid even for these small gaps. Still, note that the spin signal given by the 3D simulation is always slightly lower than the 1D one, which means that the effective polarization extracted using the one dimensional model has to be considered as a lower bound.

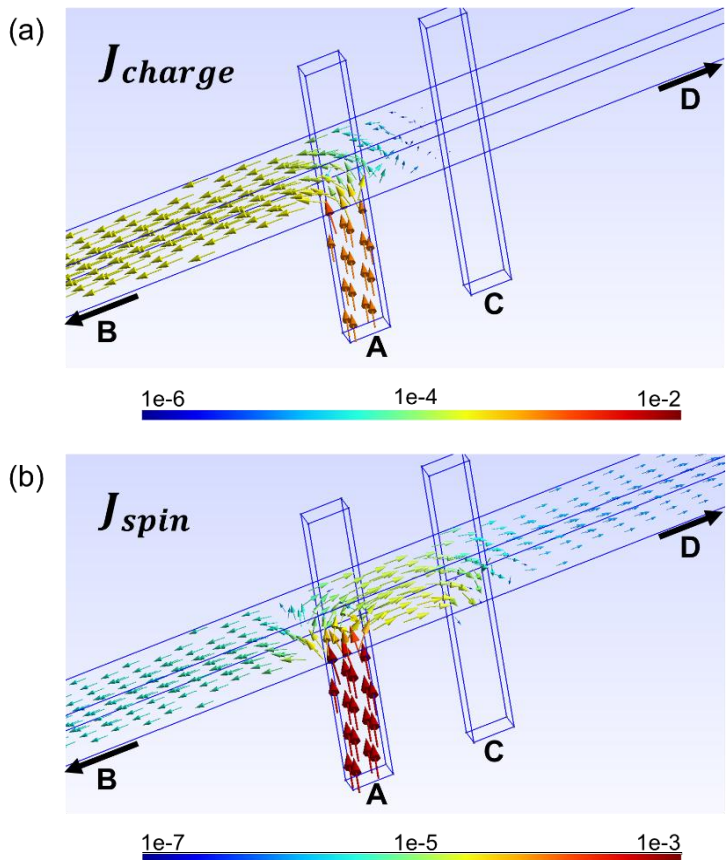

**Figure 4**
a) Display of the charge current densities simulated by FEM in a 150nm gap LSV geometry. The lattice resolution is 1,5 nm long. The parameters material are identical to the ones from the 1D model. One ampere his injected from A to B by using Neumann conditions. Magnetizations of both electrodes are identical, along the easy axis. The detected voltage is the differences of the potential between C and D edges. Spin signal correspond to the difference of this voltage between parallel and antiparallel states. b) Spin current distribution of the same simulation.

To correctly support the comparison between NiFe and CoFe, it should be pointed out that our spin signal values with NiFe LSVs are relevant when compared to the literature. Indeed, the spin signal amplitude of our 150 nm gap NiFe-Cu LSV reaches 9.1 mΩ at 300K and 22.0 mΩ at 77 K, with an effective polarization of NiFe at 77K around 0.3-0.35 [20]. These spin signal amplitudes match or exceed signals currently recorded in NiFe-based transparent LSVs [3,4,13].

V. SUMMARY

To conclude, we have demonstrated that CoFe-based transparent LSVs can exhibit spin signal amplitudes up to ten times higher than NiFe-based identical devices. Spin signals as large as 50 mΩ and 80 mΩ have been obtained at 300K and 10 K respectively. We have used a one dimensional model to extract the parameters of CoFe, and explained this increase of signal by a higher effective polarization. We have also performed finite element method simulations to verify the validity of the 1D model. In the light of this study, the CoFe alloy is a promising candidate for the replacement of Permalloy in lateral spintronic devices.

ACKNOWLEDGMENTS

This work was partly supported by the French Agence Nationale de la Recherche (ANR) through Projects SPINHALL (2010-2013) and SOSPIN (2013-2016).